\documentclass[%
 preprint,
nofootinbib,
 amsmath,amssymb,
 aps,
pra,
 longbibliography,
 lengthcheck,%
]{revtex4-1}

\usepackage{graphicx}
\usepackage{dcolumn}
\usepackage{bm}
\usepackage{hyperref}

\def\ba{\begin{equation}\begin{array}{c}}
\def\ea{\end{array}\end{equation}}
\def\be{\ba\displaystyle}
\def\ee{\ea}

\newcommand{\ra}{\rangle}
\newcommand{\la}{\langle}

\newcommand{\sign}{{\rm sign}}
\newcommand{\Ci}{{\rm Ci}}

\renewcommand{\H}{\hat H}
\newcommand{\kF}{k_{\rm F}}

\newcommand{\vs}{v_{\rm s}}

\renewcommand{\Pi}{\hat P}
\newcommand{\Ptot}{\hat P_{\rm tot}}

\begin{document}

\preprint{APS/123-QED}

\title{
Necessary and sufficient condition for quantum adiabaticity \\
in a driven one-dimensional impurity-fluid system
}

\author{Oleg Lychkovskiy$^{1,2}$}
\author{Oleksandr Gamayun$^{3}$}
\author{Vadim Cheianov$^{4}$}

\affiliation{$^1$ Skolkovo Institute of Science and Technology,
Skolkovo Innovation Center 3, Moscow  143026, Russia}
\affiliation{$^2$ Steklov Mathematical Institute of Russian Academy of Sciences,
8 Gubkina St., Moscow 119991, Russia,}
\affiliation{$^3$
Institute for Theoretical Physics, Universiteit van Amsterdam,
Science Park 904, Postbus 94485, 1098 XH Amsterdam, The Netherlands}
\affiliation{$^4$ Instituut-Lorentz, Universiteit Leiden,
P.O. Box 9506, 2300 RA Leiden, The Netherlands}


\date{\today}

\begin{abstract}
We study under what conditions the quantum adiabaticity is maintained in a closed many-body system consisting of a one-dimensional fluid and an impurity particle dragged through the latter by an external force. We employ an effective theory describing the low-energy sector of the system to derive the time dependence of the adiabaticity figure of merit -- the adiabatic fidelity. We find that in order to maintain adiabaticity in a large system the external force, $F_N$, should vanish with the system size, $N$, as $1/N$ or faster. This improves the necessary adiabatic condition $F_N=O(1/\log N)$ obtained for this system earlier \cite{lychkovskiy2018quantum}. Experimental implications of this result and its relation to the quasi-Bloch oscillations of the impurity are discussed.


\end{abstract}

\pacs{Valid PACS appear here}
\maketitle


\section{Introduction}

Quantum adiabatic theorem is a fundamentally important result in the theory of quantum systems with time-dependant Hamiltonians. In essence, it states that a system initially prepared in an instantaneous eigenstate of a Hamiltonian remains arbitrarily close to the (time-evolving) instantaneous eigenstate provided the ramp rate (i.e. the rate of change of the Hamiltonian) is slow enough \cite{born1926,born1928beweis}. When it comes to applying the adiabatic theorem in practice, the key question to be addressed is how slow  "slow enough" is. While this question can be exhaustively answered for a simple two-level system \cite{landau1932theorie,zener1932non}, it becomes complicated for many-body systems and/or for continuous quantum systems with infinite-dimensional Hilbert spaces. Although numerous sufficient conditions for adiabaticity are known (see the pioneering work \cite{kato1950} and the review \cite{albash2018adiabatic}), they often prove to be inapplicable for continuous quantum systems due to the divergence of operator norms entering these conditions. Recently a necessary condition for adiabaticity have been proven \cite{lychkovskiy2017time} which is free from this shortcoming and is well-suited for applying to many-body systems. Anyway, any adiabatic condition, whether sufficient or necessary, provides only a bound on the driving rate, without an indication how tight this bound  is.\footnote{One could imagine that both sufficient and necessary condition are available and provide bounds which are close to each other. In practice, however, such a fortunate occurrences are rare if not extinct, at least in the many-body context.}

In the present paper we derive a necessary and sufficient condition for quantum adiabaticity in a many-body system consisting of a one-dimensional fluid and an impurity particle dragged through the latter by a constant external force. This system exhibits a spectacular phenomenon predicted in refs. \cite{Gangardt2009,schecter2012dynamics,schecter2012critical} and experimentally observed in ref. \cite{meinert2016bloch} -- quasi-Bloch oscillations of the impurity's velocity and position. These oscillations are somewhat reminiscent to the Bloch oscillations of a single particle in a periodic potential. Their root cause is the nontrivial spectral edge of the impurity-fluid system in one dimension, which is periodic in the thermodynamic limit as a function of the total momentum, with the period $2\pi \rho$ determined by the density of the fluid $\rho\equiv N/L$ (here $N$ is the number of particles of the fluid, $L$ is the linear dimension of the system, and we set  $\hbar=1$ throughout the paper).   There are two important features, however, distinguishing quasi-Bloch oscillations from the conventional Bloch oscillations. First, intriguingly, quasi-Bloch oscillations occur in the translation-invariant system, in the absence of any external periodic potential. Second, quasi-Bloch oscillations are a genuinely many-body phenomenon.

The exact conditions for the occurrence of the quasi-Bloch oscillations is a matter of a controversy \cite{Gangardt2009,schecter2012dynamics,schecter2012critical,Gamayun2014kinetic,Gamayun2014keldysh,lychkovskiy2015perpetual,schecter2014comment,Gamayun2014reply,schecter2016quantum}. However, it is undisputable that the many-body adiabaticity is a sufficient (though, in general, not necessary) condition for the quasi-Bloch oscillations \cite{schecter2012dynamics}. This is the reason for our interest in conditions for adiabaticity in the one-dimensional impurity-fluid system.

Recently we have applied a {\it necessary} adiabatic condition of ref. \cite{lychkovskiy2017time}  to the impurity-fluid system, with the result that  in order to maintain adiabaticity the driving force $F_N$ should vanish with the system size (the density of the fluid being fixed) {\it at least} as fast as $O(1/\log N)$~\cite{lychkovskiy2018quantum}. This result demonstrates that the adiabaticity does not survive in the thermodynamic limit, as is expected on general grounds for a gapless many-body system~\cite{balian2007microphysics}. However, if the $O(1/\log N)$ scaling were a true scaling of the maximal force tolerated by adiabaticity, it would be well possible to observe the adiabatic evolution in state-of-the-art cold atom experimental settings with moderately large $N$, e.g. with $N\sim 100$ like in the experiment of ref. \cite{meinert2016bloch}. This observation motivated us to search for a {\it necessary and sufficient} adiabatic condition in the impurity-fluid system.

Here we report such a condition obtained in the framework of an effective theory describing an impurity slowly moving in a 1D quantum fluid  \cite{tsukamoto1998critical,kamenev2009dynamics,zvonarev2009edge}. The condition has the form $F_N < O(1/N)$,  which is a dramatic quantitative difference from the logarithmic scaling obtained previously. This result implies that maintaining many-body adiabaticity in the impurity-fluid system is a very challenging experimental task.




The paper is organized as follows. After a general discussion of the notion of adiabaticity in Sec. \ref{sec figure of merit}, we introduce the impurity-fluid  system and its effective description in Sec. \ref{sec impurity-fluid system}. The diagonalization of the effective Hamiltonian at a given moment of time is reviewed in Sec. \ref{sec gs}. The solution of the full dynamical problem is presented in Sec. \ref{sec dynamics}. In Sec. \ref{sec results} the results are presented and their immediate experimental implications are discussed. In Sec. \ref{sec discussion} we summarize our results and make a couple of concluding remarks. Most of the technicalities  are reserved to the appendices.




\section{Adiabaticity: Figure of merit\label{sec figure of merit}}
We start from introducing the notion of adiabaticity in quantitative terms. Consider a parameter-dependent Hamiltonian $H_Q$, $Q$ being for a moment an abstract parameter. We introduce time dependence in this Hamiltonian by assuming that $Q$ linearly varies with time $t$,
$$Q=F_N t.$$
At the moment $F_N$ is treated merely as another abstract parameter quantifying the dirivng rate. The subscript $N$ in $F_N$ indicates that the driving rate may, in general, scale with the system size.  For each $Q$ one can define an instantaneous ground state, $\Phi_Q$,
which is the lowest eigenvalue solution to the Schr\"odinger's stationary equation,
\begin{equation}\label{eigenproblem}
\H_Q\, \Phi_Q = E_Q \, \Phi_Q.
\end{equation}
Here $E_Q$ is the instantaneous ground state energy. We assume that the ground state is non-degenerate for any~$Q$.

The dynamics of the system is governed by the Schr\"odinger  equation, which can be written in a rescaled form as
\begin{equation}\label{Schrodinger equation}
\left(\H_{Q}\,-\,i\,F_N\, \partial_Q\right) \Psi_Q = 0.
\end{equation}
Here  $\Psi_Q$ is the state vector of the system which depends on time through the time-dependent parameter $Q$.  Initially, at $t=0$ (or, equivalently, at $Q=0$), the system is prepared in the instantaneous ground state:
\begin{equation}\label{initial condition}
 \Psi_0= \Phi_0.
\end{equation}

The evolution is called adiabatic as long as the state of the system, $\Psi_Q$, stays close to the instantaneous ground state,  $\Phi_Q$.
To what extent adiabaticity is preserved during the evolution is quantified by the {\it adiabatic fidelity} $\mathcal F_Q$, which is the probability that the evolved state coincides with the instantaneous ground state,
\be\label{w}
\mathcal F_Q\equiv\left|\la \Phi_Q \,|\, \Psi_Q \ra \right|^2.
\ee
Perfect adiabaticity would imply $\mathcal F_Q =1$.
Calculating $\mathcal F_Q$ for the driven impurity-fluid system is the main goal of our study.

Another useful quantity is the {\it adiabatic mean free path}  $Q_*$ which quantifies how far can the system travel in the parameter space for a given driving rate before the adiabaticity breaks down. We define this quantity as the smallest positive solution of the equation
\be\label{mean free pass}
\log \mathcal F_{Q_*}=-1.
\ee

\section{Impurity-fluid system\label{sec impurity-fluid system}}

\subsection{Preliminary considerations}

An object of our study is a one dimensional many-body system, consisting of a quantum fluid and an impurity particle. A constant external force $F_N$ is exerted upon the impurity. The fluid consists of $N$ identical particles, either fermions or bosons. The particles of the fluid interact with the impurity and, in general, with each other.

As a preliminary step, we discuss how to describe a one-body problem of a {\it noninteracting} impurity particle pulled by an external force. This can be done conveniently by introducing a time-dependent Hamiltonian
\be\label{H imp}
\H^{\rm imp}_Q=\frac{(\Pi+Q)^2}{2m},
\ee
where $m$ is the mass of the impurity, $\Pi\equiv -i \partial/\partial X$ is the canonical momentum of the impurity and $X$ is the coordinate of the impurity. Periodic boundary conditions with some period $L$ are implied. In this context $Q=F_N t$ is the impulse of the force.



The interacting impurity-fluid system is described by the microscopic Hamiltonian
\begin{align}\label{H microscopic}
\H_Q^{\rm micro}= \H^{\rm imp}_Q + \H^{\rm f}+\hat H^{\rm if},
\end{align}
where $\H^{\rm f}$ is the Hamiltonian of the fluid which includes the kinetic term and the pairwise interactions between the particles of the fluid, and $\hat H^{\rm if}$ is the impurity-fluid coupling. We do not specify microscopic Hamiltonians $\H^{\rm f}$ and $\hat H^{\rm if}$ explicitely since our analysis will be based on an effective low-energy model described in the next section.

A general feature of translation-invariant one-dimensional systems described by the Hamiltonian \eqref{H microscopic} is that eigenenergy as a function of $Q$ is periodic in thermodynamic limit \cite{Giamarchi2003}. The period is determined by the number density of particles and, in our case, is given by $2\kF$ up to finite size corrections, where $\kF\equiv \pi N/L = \pi \rho$. The latter quantity sets the typical momentum scale of the problem. For the fluid consisting of noninteracting fermions, $\kF$ coincides with the Fermi momentum. It should be emphasized, however, that in general case we do not ascribe any ``fermionic'' meaning to $\kF$. In particular, we consider bosonic and fermionic fluids on equal footing. Note that Fermi statistics plays no role in the above-mentioned periodicity of eigenenergies, which is present for bosons as well.


\subsection{Effective Hamiltonian}

Under fairly general conditions the low-lying excitations of a one-dimensional quantum fluid can be treated by means of an effective Luttinger liquid theory \cite{haldane1981luttinger}. This theory can be extended to describe the low-energy sector of the one-dimensional impurity-fluid system  \eqref{H microscopic} \cite{tsukamoto1998critical,kamenev2009dynamics,zvonarev2009edge,imambekov2012one}. This extension is valid for sufficiently small absolute value of the velocity of the impurity, $v_Q$ (below we will discuss this condition in more detail).
The corresponding effective Hamiltonian reads \cite{tsukamoto1998critical,kamenev2009dynamics,zvonarev2009edge,imambekov2012one}
\begin{align}\label{H(Q)}
\H = & v_Q\, \Pi + \vs \sum_q |q|\hat a_q^\dagger  \hat a_q -  \nonumber\\
  & \frac{1}{\sqrt{2\pi L}} \sum_q \sqrt{q}\, \delta_Q^q \,\, \Delta v_Q^q\, \,(\hat a_q^\dagger e^{-i q X} +\hat a_q e^{i q X} ).
\end{align}
Here $v_s$ is the sound velocity of the fluid, $a^\dagger_q,\,\hat a_q$ are creation and annihilation operators of bosonic excitations of the fluid carrying momentum $q$,
\be
\Delta v_Q^q\equiv \vs-v_Q \,\sign q
\ee
and\footnote{Observe that the definition of $v_Q$ is self-consistent in the framework of the model \eqref{H(Q)}:
$
\hat V \equiv i[\H,X]=v_Q,
$
i.e. the operator of impurity's velocity,  $\hat V$, is equal to the time-dependent $c$-number~$v_Q$.}
\be
\delta_Q^q=\left\{
\begin{array}{ll}
\delta_Q^+, & q>0,\\
0, & q=0 \\
\delta_Q^{\rm -}, & q<0
\end{array}
\right.
\ee
is the scattering phase which is determined by the impurity-fluid interaction.\footnote{\label{footnote}Strictly speaking, the boson operators $a_q$, $a_q^\dagger$ might also depend on $Q$. This subtle issue is a particular instance of a general problem of ambiguity of a connection on the bundle of Hilbert spaces over a space of external parameters which vary in time (see e.g. ref. \cite{moore2017comment} for a discussion). Following an established practice \cite{tsukamoto1998critical,kamenev2009dynamics,zvonarev2009edge,imambekov2012one}, we ignore this possible dependence. This is to say, we assume that this dependence is either absent or produces corrections which are subleading for considered ranges of $Q$ (see below). This assumption is supported by an independent analysis within a microscopic integrable model, see Appendix~\ref{appendix BA}.
%
}
%


In eq. \eqref{H(Q)} and throughout the paper the increment in the sums over $q$ equals the momentum quantum $\delta k\equiv 2\pi/L$. For definitness, we employ the ultraviolet cutoff equal to $\kF$, although the exact cutoff value does not enter the final results.


The total {\it canonical} momentum
\be
\Ptot \equiv \Pi+\sum_q q \, \hat a_q^\dagger \hat a_q
\ee
commutes with the Hamiltonian $H$ and therefore is conserved. Its eigenvalue is quantized in units of $\delta k$. The canonical momentum should not be confused with the total {\it kinetic} momentum, $\hat P_{{\rm kin}}=\Ptot+Q$, which is not quantized and grows linearly with time due to the action of the external force.

In fact, the effective model \eqref{H(Q)} is well-defined only in a subspace of the Hilbert space corresponding to some  eigenvalue  $P_{\rm tot}$ of the total canonical momentum. In this subspace the kinetic momentum also has a well-defined, though varying with time, eigenvalue $P_{{\rm kin}}$. Effective ``constants'' $v_Q$, $\delta_Q^+$ and $\delta_Q^{\rm -}$  are in fact functions of the kinetic momentum, $P_{{\rm kin}}=P_{\rm tot}+Q$ (and, therefore,  notations $v_{P_{{\rm kin}}}$, $\delta_{P_{{\rm kin}}}^+$ and $\delta_{P_{{\rm kin}}}^{\rm -}$ would be more consistent).  We, however, choose to fix $P_{\rm tot}$ and refrain from referring to it explicitly throughout the paper, except the present section and Appendix \ref{appendix diagonalization}. The only importance of the precise value of  $P_{\rm tot}$ is to fix $v_Q$, $\delta_Q^+$ and $\delta_Q^{\rm -}$ at $Q=0$.


The range of validity of the effective Hamiltonian \eqref{H(Q)} is a somewhat subtle issue. This Hamiltonian is designed to describe a low-energy sector of the Hilbert space, i.e. an energy shell of a width $\Delta E_{P_{{\rm kin}}}$ above the ground state. The subscript in $\Delta E_{P_{{\rm kin}}}$ indicates that the range of validity of the effective model varies with $P_{{\rm kin}}$.  In certain cases $\Delta E_{P_{{\rm kin}}}$ is nonzero in the whole Brillouin zone, $-\kF<P_{{\rm kin}}<\kF$, and vanishes only at its ends  \cite{tsukamoto1998critical,kamenev2009dynamics,zvonarev2009edge}. In particular, this is the case for an integrable model solved by McGuire \cite{mcguire1965interacting} which is discussed in Appendix \ref{appendix BA}.  In other cases, however, the effective model \eqref{H(Q)} breaks down in a finite portion of the Brillouin zone. In particular, this happens for a sufficiently light impurity weakly interacting with a one-dimensional fluid \cite{Gamayun2014kinetic,lychkovskiy2015perpetual}. In general, one expects that $\Delta E_{P_{{\rm kin}}}$ is nonzero as long as the impurity moves with the velocity below the generalized critical velocity $v_{\rm c}\leq \vs$ which ensures absence of the Cherenkov-like radiation \cite{lychkovskiy2015perpetual}. The latter critical velocity is typically on the order of $\vs$. To summarize, to be on the safe side, one can assume
\be\label{validity range}
v_Q\ll \vs,
\ee
although the actual range of validity of the effective Hamiltonian \eqref{H(Q)} can be much wider.






\section{Instantaneous ground state\label{sec gs}}

The first ingredient required for calculating the adiabatic fidelity $\mathcal F_Q$ is the instantaneous ground state $\Phi_Q$ of the Hamiltonian \eqref{H(Q)}. The latter can be diagonalized exactly \cite{tsukamoto1998critical,kamenev2009dynamics,zvonarev2009edge}. We describe and discuss the diagonalization procedure and identification of the ground state in the Appendix \ref{appendix diagonalization}. Here we give the final result which reads
\be\label{diagonalization}
e^{\hat W_Q}\H_Qe^{-\hat W_Q} = \H^{\rm d}_Q+C,
\ee
where
\be\label{H diag}
\H^{\rm d}_Q=v_Q \hat{P} + \vs \sum_q |q|\hat a_q^\dagger a_q ,
\ee
and $C$ is a $c$-number  which is omitted in what follows.\footnote{In fact, a $Q$-dependent $c$-number responsible for reproducing the correct ground state energy is already omitted in the definition \eqref{H(Q)} of $H$. Predicting the ground state energy is beyong the scope of the effective low-energy model.} The anti-Hermitian operator $W$ reads
\be\label{W_Q}
\hat W_Q=\sum_q (\alpha^q_Q \, \hat a_q^\dagger e^{-i q X} - \overline{\alpha_Q^q} \, \hat a_q \, e^{i q X}),
\ee
where
\be\label{alpha}
\alpha^q_Q=-\frac{\delta_Q^q}{\sqrt{2\pi L |q|}}
\ee
and the overbar in $\overline{\alpha_Q^q}$ and elswhere refers to the complex conjugation.

The ground state of $\H_Q$ for a fixed total canonical momentum $P_{\rm tot}$ reads
\be\label{evolution of gs}
\Phi_Q = e^{-\hat W_Q}|{\rm vac}, \,  P_{\rm tot} \ra,
\ee
where $|{\rm vac}, \,  P_{\rm tot} \ra\equiv |{\rm vac}\ra \otimes |  P_{\rm tot} \ra$ is a product state with $|{\rm vac}\ra$ being a Fock vacuum with respect to bosonic operators $\hat a_q$ and $| P_{\rm tot} \ra$ being the state of impurity with the momentum $P_{\rm tot}$.



\section{Dynamics\label{sec dynamics}}
\subsection{Dynamical diagonalization}

The second ingredient  for calculating  $\mathcal F_Q$ is the dynamical state vector $\Psi_Q$ evolving according to the Schr\"odinger  equation \eqref{Schrodinger equation}. Remarkably, the dynamics described by this equation is integrable, in the sense that the operator
$\left( \H_{Q}-i\, F \, \partial_Q\right)$ can be diagonalized by a unitary transformation analogous to the transformation~\eqref{diagonalization}:
\be\label{diagonalization dynamical}
e^{\hat Y_Q} \left( \H_{Q}-i\, F \, \partial_Q\right)  e^{-\hat Y_Q} = \H^{\rm d}_Q-i\, F \, \partial_Q+C'.
\ee
Here $C'$ is a $c$-number which will be omitted in what follows, 
\be\label{Y_Q}
\hat Y_Q = \sum_q (\beta_Q^q \hat a_q^\dagger e^{-i q X} - \overline{\beta_Q^q} \hat a_q\,e^{i q X})
\ee
and the coefficients $\beta_Q^q $ satisfy the differential equation
\be\label{differential eq}
i F\partial_Q \beta_Q^q  - |q| \Delta v_Q^q\beta_Q^q  -  \frac{\delta_Q^q }{\sqrt{2\pi L}} \sqrt{|q|} \Delta v_Q^q= 0.
\ee
The key insight behind the dynamical diagonalization is that $[ Y_Q,\partial_Q Y_Q]$ is a $c$-number, and therefore
\be\label{derivative}
-i\, F \, e^{\hat Y_Q}\,\left(   \partial_Q  e^{-\hat Y_Q}\right) = i\, F \,(\partial_Q Y_Q +\frac12 [ Y_Q,\partial_Q Y_Q])
\ee
is linear in boson operators and has the same structure as $Y_Q$ and interaction term of the Hamiltonian \eqref{H(Q)}.

In order to satisfy the initial condition of the  Schr\"odinger  equation \eqref{initial condition}  we supplement the differential equation \eqref{differential eq}  with the initial condition
\be
\beta_0^q =\alpha^q_0.
\ee

Eq. \eqref{differential eq} can be solved in quadratures, with the result
\be\label{beta exact}
\beta_Q^q =\alpha_Q^q  - \int_0^Q \partial_{Q'}\alpha^q_{Q'}\,
\exp\left(-i \frac{ |q|}{F_N}\int_{Q'}^Q d \widetilde Q \, \Delta v^q_{\widetilde Q} \right)\,d Q'
\ee

Eq. \eqref{diagonalization dynamical} entails that  the dynamical state of the system evolves according to
\be\label{evolution of state}
\Psi_Q = e^{-\hat Y_Q }|{\rm vac},\,  P_{\rm tot} \ra.
\ee
where an irrelevant $c$-number phase factor is omitted.




\subsection{Dynamics of adiabatic fidelity}

With $\Phi_Q$ and $\Psi_Q$ in hand, we are prepared to proceed to calculating the adiabatic fidelity. Substituting  eqs. \eqref{evolution of gs} and \eqref{evolution of state} into the definition  \eqref{w} and applying the Baker-Campbell-Hausdorff formula, we obtain
\be\label{w prefinal}
\mathcal F_Q =   \exp\left( - \sum_q  |\beta_Q^q-\alpha_Q^q|^2 \right).
\ee




\section{Results\label{sec results}}

\subsection{Adiabatic fidelity  and adiabatic mean free path\label{subsec results}}

In principle, eqs. \eqref{w prefinal} along with eqs. \eqref{alpha} and \eqref{beta exact} allows one to calculate ${\cal F}_Q$  for any values of parameters within the validity range of the model  \eqref{H(Q)}. However, a further asymptotic analysis is required to reveal the scaling properties of $\mathcal F (Q)$. Here we present the main results of such an analysis, referring the reader to Appendix \ref{appendix asymptotic analysis} for details.





The asymptotic behavior of the adiabatic fidelity in the limit of large system size depends on how the force $F_N$ scales with $N$. Before turning to a general case, we consider an important special case of the force independent from the system size.
In this case we obtain
\begin{align}\label{result 0}
\log{\cal F}_Q= &\, -\xi^2\,\left(\frac{Q}{\kF}\right)^2 \, \log N
\end{align}
with
\begin{align}\label{xi}
\xi^2 = &
\,\kF^2\,
\left.
\left(
\left(\frac{\partial_Q \delta_Q^+}{2\pi} \right)^2+\left(\frac{\partial_Q \delta_Q^-  }{2\pi} \right)^2
 \right)
 \right|_{Q=0}
\end{align}
This is the leading term of the double asymptotic expansion of $\log{\cal F}_Q$ in the limit of
\begin{align}
N& \gg  1,  & N/L {\rm ~~~fixed}, \label{limit N}\\
Q& \ll \kF, & \label{limit Q}
\end{align}
where the limit \ref{limit N} is taken first.

Quite remarkably, the force does not enter eq. \eqref{result 0}. In fact, the adiabatic fidelity follows the orthogonality catastrophe overlap, ${\cal F}_Q \simeq  {\cal C}_Q \equiv \left| \langle \Phi_0|\Phi_Q \rangle \right|^2$, in accordance with the scenario described in ref.~\cite{lychkovskiy2017time}. One can obtain eq. \eqref{result 0} without solving the Schr\"odinger  equation~\eqref{Schrodinger equation} by calculating the orthogonality overlap $\left| \langle \Phi_0|\Phi_Q \rangle \right|^2$ and applying the general result of ref.~\cite{lychkovskiy2017time}, as detailed in Appendix \ref{appendix OC and A}. However, this method is applicable in a superficially narrow range of $Q$ which shrinks when $N$ grows, $Q=o(1)$. Solving the dynamical problem shows that the validity range of the approximation  $\cal F_Q \simeq  \cal C_Q$ appears to be larger than that guaranteed by the rigorous result of ref.~\cite{lychkovskiy2017time} -- analogous conclusions have been made on the basis of explicit solutions of other models \cite{lychkovskiy2017time}.





Now we turn to a more general case when the force $F_N$ can vanish with the system size, but not faster than $1/N$. In this case one needs to consider separately two ranges of $Q$:
\begin{align}\label{result 1}
\log{\cal F}_Q=
-
&\xi^2\,\left(\frac{Q}{\kF}\right)^2\,
 \nonumber \\
& \times
\left\{
\arraycolsep=1.4pt\def\arraystretch{2}
\begin{array}{ll}
  \log N, &   Q\ll \frac{F_N}{\kF \vs},\\
\log\left(N\,\frac{F_N }{\kF^2 \vs }\right), &     \frac{F_N}{\kF \vs}\ll Q \ll \kF,
\end{array}
\right.
\end{align}
where $\xi$ is given by eq. \eqref{xi}.

One can see from the first line of eq. \eqref{result 1} that for small momenta the result coincides with those of  eq. \eqref{result 0} and, in fact, again can be obtained by the method of ref. \cite{lychkovskiy2017time} without considering dynamics, see Appendix \ref{appendix OC and A}.
Solving  the time-dependent Schr\"odinger  equation explicitly is mandatory for obtaining the expression for larger momenta (the second line of eq. \eqref{result 1}). Observe that the latter expression manifestly depends on the force.



One can easily find the adiabatic mean free path, $Q_*$, from eq. \eqref{result 1}:
\begin{align}\label{result 2}
\frac{Q_*}{\kF}=&
\frac1\xi
\left\{
\arraycolsep=3pt\def\arraystretch{2}
\begin{array}{ll}
   \left( \log N\right)^{-\frac12},   &  \frac{F_N}{\kF^2 \vs }   \gg \frac1{\sqrt{\log N}}, \\
  \left( \log\frac{N \,F_N}{\kF^2 \vs }\right)^{-\frac12}, & \frac1N\ll \frac{F_N}{\kF^2 \vs }  \ll \frac1{\sqrt{\log N}}.
\end{array}
\right.
\end{align}
Again, solving the time-dependent Schr\"odinger  equation is essential for obtaining the second line of this expression, while the first line is obtained in ref. \cite{lychkovskiy2018quantum} without considering dynamics.


Eq. \eqref{result 2} implies that
\be\label{result 3}
F_N=O(1/N)
\ee
is a necessary asymptotic condition for the many-body adiabaticity. While the scaling $F_N=O(1/N)$ is beyond the range of validity of eq. \eqref{result 2}, the second line of this equation indicates that the condition \eqref{result 3} can be also sufficient for adiabaticity. This is indeed the case, as is proven in Appendix \ref{appendix sufficient}.  Thus, we establish eq. \eqref{result 3} as a necessary and sufficient condition for adiabaticity in the driven impurity-fluid system in the large system size limit.

At this point it is worth to comment on the results of ref. \cite{lychkovskiy2018quantum} where we have studied adiabaticity breakdown in the integrable impurity-fluid system by the method of ref.~\cite{lychkovskiy2017time}. Integrability allowed us to calculate ${\cal C}_Q$ explicitly and establish a necessary adiabatic condition $F_N\leq O(1/\log N)$. Obviously, the result \eqref{result 3} of the present paper is much stronger. However, the analysis of a microscopic integrable model retains its value since it underpins the effective model \eqref{H(Q)} and illustrates how the phenomenological parameters of the effective model are related to microscopic parameters. We present this analysis in Appendix \ref{appendix BA}. A particularly interesting conclusion from this analysis is that the result of calculation of  ${\cal C}_Q$  within the integrable model and within the effective model  \eqref{H(Q)} coincide for all $Q$, not just within the conservative validity range \eqref{validity range}. This indicates that the actual validity range of the effective treatment can be larger.


\subsection{Experimental implication}

Damped oscillations of an impurity particle driven through a 1D quantum fluid were observed in a recent experiment \cite{meinert2016bloch}, where a quantum fluid consisted of $N\simeq60$ Cesium atoms and the force was equal to $1/3$ of the gravitational force. In this experiment the adiabaticity faded away on a time scale smaller than one period of oscillations. This was clear from the direct numerical simulation of the experimental setting which showed the increase of the energy of the system above the ground state energy already during the first half-period of oscillations   \cite{meinert2016bloch}. We confirm this conclusion by calculating the adiabatic mean free path according to eq. \eqref{result 2}, which appears to be around $\kF$ throughout the whole range of experimental conditions.

One may wonder whether it is possible to maintain the many-body adiabaticity for at least a few cycles of oscillations by e.g. applying a smaller force. Our analysis indicates that this can be challenging. The reason is that in practice the spatial amplitude of oscillations is limited by the size of  the  quasi-1D optical cigar-shaped trap. It is easy to see that this amplitude is inversely proportional to the force. This constrains the force to satisfy
$
F_N \gtrsim \kF^3/(2\pi m_* N),
$
where $m_*$ is the effective mass of the impurity in the fluid.
This can be only marginally consistent with the adiabatic condition \eqref{result 3}. Therefore maintaining adiabaticity for several periods of oscillations would require an extremely careful choice of the driving force -- not too high to sustain adiabaticity but not too low to keep the spatial amplitude of oscillations within the trap size.

\section{Summary and concluding remarks\label{sec discussion}}

To summarize, we have analyzed the dynamics of the many-body adiabatic fidelity, ${\cal F}_Q$, in a one-dimensional impurity-fluid system where a force applied to the impurity pulls the latter through the fluid. We have employed an effective low-energy theory which enabled us to find explicit expressions for  ${\cal F}_Q$ and the adiabatic mean free path, $Q_*$, in terms of the size of the system and the effective parameters of the fluid and the impurity-fluid coupling.
Our results imply that
the state vector of the impurity-fluid system completely departs from the instantaneous ground state already for acquired momenta which are vanishingly small in thermodynamic limit,
unless  the force scales down with the system size as $1/N$ or faster. This dramatically improves the necessary adiabatic condition $F_N\leq O(1/\log N)$ obtained previously~\cite{lychkovskiy2018quantum}.

It is remarkable that quantum adiabaticity breaks down already at small acquired momenta. When the acquired momentum reaches the vicinity of $\pi \rho$, our quantitative results may be inapplicable since the vicinity of $Q=\pi \rho$ can be beyond the range of validity of the effective model employed. However, the adiabaticity is anyway gone at this point (unless $F_N\leq O(1/N)$). It is worth noting that crossing the $Q=\pi \rho$ point is potentially the most dangerous part of the oscillation cycle with respect to preserving adiabaticity \cite{schecter2014comment,Gamayun2014reply}. This means that the adiabatic condition can become only more stringent beyond the range of validity of our approach.

It should be emphasized that while the many-body adiabaticity would be sufficient to ensure quasi-Bloch oscillations in an arbitrary  one-dimensional impurity-fluid system, adiabaticity breakdown is not necessarily fatal for oscillations. This has been confirmed in the experiment of ref. \cite{meinert2016bloch}. On the theoretical side, there is a consensus that quasi-Bloch oscillation of a sufficiently heavy impurity can occur in the thermodynamic limit for a force which is finite in the thermodynamic limit but sufficiently small as compared to other {\it intensive} (i.e. independent on the system size) quantities of the system \cite{Gangardt2009,schecter2012dynamics,schecter2012critical,Gamayun2014kinetic,lychkovskiy2015perpetual}. The latter condition defines a notion of a {\it thermodynamic} adiabaticity, as contrasted to the genuine quantum many-body adiabaticity studied in the present paper.\footnote
{An instructive demonstration of the subtlety of interplay between the two concepts was recently presented in ref. \cite{gamayun2018impact}, where an impurity-fluid system with a time-dependent coupling constant (but in the absence of a force) was considered. It appeared that the outcomes of the thermodynamically adiabatic and the genuinely quantum-adiabatic evolutions could be identical or dramatically different in the same system, depending on the choice of the initial state.
}
It is a matter of a debate whether thermodynamic adiabaticity is sufficient for quasi-Bloch oscillations of a light impurity \cite{Gangardt2009,schecter2012dynamics,schecter2012critical,Gamayun2014kinetic,Gamayun2014keldysh,lychkovskiy2015perpetual,schecter2014comment,Gamayun2014reply,schecter2016quantum}. The present paper does not contribute to this debate.

\begin{acknowledgements}
We thank M. Zvonarev for useful discussions. The work was supported by the Russian Science Foundation under the grant N$^{\rm o}$ 17-71-20158.
\end{acknowledgements}

\appendix

\section{Diagonalization for a fixed $Q$ \label{appendix diagonalization}}

Here we present details on diagonalization of the Hamiltonian $H(Q)$ at a given $Q$.

\subsection{Unitary transformation}
Unitary transformation generated by $W$ defined by eq. \eqref{W_Q} with arbitrary coefficients $\alpha^q_Q$
yields
\be
e^{\hat W_Q} \hat a_q e^{-\hat W_Q}= \hat a_q - \alpha^q_Q  e^{-i q X}
\ee
%
%
and
\be \label{Pi transformed}
e^{\hat W_Q} \Pi e^{-\hat W_Q}=\Pi+ \sum_q q (\alpha^q_Q \,  \hat a_q^\dagger e^{-i q X} + h.c.)-\Delta P,
\ee
where the constant $\Delta P$ reads
\be\label{Delta P}
\Delta P=\sum_q q |\alpha^q_Q|^2.
\ee
As a consequence,
\be\label{Ptot transformed}
e^{\hat W_Q} \Ptot e^{-\hat W_Q}=\Ptot
\ee
and
\begin{align}\label{H transformed}
& e^{\hat W_Q} \H e^{-\hat W_Q}  =  \H_{\rm d} + C -  \\
 &  - \sum_q |q|
 \left(
 (\alpha^q_Q \Delta v^q+ \frac{\delta_Q^q}{\sqrt{2\pi L |q|}}\Delta v_Q^q)\,  \hat a_q^\dagger e^{-i q X}
 +h.c.
\right)\nonumber
\end{align}
with $\H_{\rm d}$ given by eq. \eqref{H diag} and the constant $C$ given by
\be\label{C}
C= \vs \sum_q |q| |\alpha^q_Q|^2 +
   \frac1{\sqrt{2\pi L}} \sum_q \sqrt{|q|}\, \delta_Q^q \Delta v_Q^q\,(\alpha^q_Q +\overline{\alpha_Q^q}).
\ee
The last term in eq. \eqref{H transformed} vanishes when one chooses coefficients $\alpha^q_Q$ according to \eqref{alpha}. In this case
\be
\Delta P = \frac{\kF}{(2\pi)^2}\left((\delta_Q^+)^2-(\delta_Q^-)^2\right).
\ee


\subsection{Ground state}

Eigenstates of the diagonalized Hamiltonian \eqref{H diag} do not depend on $Q$ and read
\be
|\{ n_q \},\, K \ra \equiv |\{ n_q \}\ra\,\otimes \,| K \ra,
\ee
where $|\{ n_q \}\ra$ is an eigenstate of the oscillator part of  \eqref{H diag} with $n_q$ bosons for each $q$, while  $| K \ra$ is the state of the impurity with momentum $K$.
The eigenvalues of the hamiltonian and the total momentum read, respectively
\begin{align}
E_{\{ n_q \}, K } & =  v_Q K + \vs \sum_q |q| \, n_q, +C \label{H eigenvalue}~~~{\rm and} \\
P_{{\rm tot}\, \{ n_q \}, K} & = K + \sum_q q \, n_q . \label{Ptot eigenvalue}
\end{align}

We wish to find the minimal eigenenergy $E_{\{ n_q \}, K }$ for a given total momentum $P_{{\rm tot}\, \{ n_q \}, K}$. To this end we introduce
\be
P_\pm\equiv \sum_{\pm q>0} |q| \, n_q.
\ee
and rewrite eqs. \eqref{H eigenvalue} and \eqref{Ptot eigenvalue}  as, respectively,
\begin{align}
E_{\{ n_q \}, P_{\rm tot}} & =  v_QK + \vs (P_+ + P_-) +C~~~{\rm and}  \\
P^{\rm tot}_{\{ n_q \}, K} & = K +P_+ - P_-.
\end{align}
This leads to
\be\label{E}
E_{\{ n_q \}, K}  =  v P_{\rm tot}+C + (\vs-v) P_+ + (\vs+v) P_-.
\ee
Since two last  terms in this equation are nonnegative while $C$ is the same for all $\{ n_q \}$ and $K$, the r.h.s. of eq. \eqref{E} has minimum at $P_+=P_-=0$. Hence the ground state of $\H_{\rm d}$ for a given value of the total momentum reads
$
|{\rm vac}, \,  P_{\rm tot}\ra
$, where $|{\rm vac}\ra$ is $|\{ n_q \}\ra$ with all $n_q=0$, and the ground state $\Phi_Q$ of $\H_Q$ is given by eq. \eqref{evolution of gs}.

We note that
\be\label{Pi average}
\la\Phi_Q| \Pi |\Phi_Q\ra = P_{\rm tot}-\Delta P.
\ee

\section{Asymptotic analysis \label{appendix asymptotic analysis}}

From eqs. \eqref{w prefinal} and \eqref{beta exact} one gets
\begin{align}\label{w preprefinal}
\log \mathcal F (Q) =     & - \int_0^Q \int_0^Q d Q' d Q'' \, \frac{\partial_{Q'} \delta_{Q'}^+}{2\pi} \,\frac{\partial_{Q''} \delta_{Q''}^+}{2\pi} \, \Sigma_+(Q''-Q') \nonumber\\
&- \Big\{
\delta_Q^+\to\delta_Q^-,\,\,\Sigma_+\to\Sigma_-
\Big\}.
\end{align}
with
\be\label{Sigma}
\Sigma_\pm \equiv \sum_{n=1}^{M-1} \frac1n   e^{-i n \phi_\pm } =-\frac12 \log(2-2\cos \phi_\pm)  +\Ci(M \phi_\pm  )
\ee
and
\be
\phi_\pm\equiv \frac{\delta k}{F_N}\int_{Q'}^{Q''} d \widetilde Q \, (\vs\mp v_Q).
\ee
Here $\Ci$ stands for the cosine integral function and we use an arbitrary momentum cutoff
\be
u \equiv (M-1) \delta k \sim \kF.
\ee
If one takes $u=\kF$, as we do in the rest of the paper, then $M=(N+1)/2$.
Note that
\be
\phi_\pm\ll 1
\ee
since we assume that $N F_N \gg 1$.

We analyze the asymptotic behavior of the r.h.s. of eq. \eqref{Sigma} in two opposite limits. The first one is
\be
M \phi_\pm   \ll 1,
\ee which is equivalent to
$Q\ll F_N/(\vs \kF)$. In this case one gets
\begin{align}\label{asymp 1}
\Sigma_\pm & =\log M + \gamma_{\rm E}+O\left((\phi_\pm M)^2 \right)+O\left(\frac1N\right) \nonumber\\
& =\log N + \gamma_{\rm E}+ \log\frac{M}{N} +O\left(\left(\frac{Q\vs\kF}{F_N}\right)^2 \right)
\end{align}
Observe that only the leading term is cutoff-independent.

The second limit is
\be
M \phi_\pm   \gg 1,
\ee
 which is equivalent to
$Q \gg F_N/(\vs \kF)$. In this case the cutoff-dependent part of the r.h.s. of eq. \eqref{Sigma} is vanishingly small,
\be
\Ci(M \phi_\pm)=O \left(\frac{1}{M \phi_\pm}\right).
\ee
We find it convenient to further expand the cutoff-independent part  of eq. \eqref{Sigma} in $\phi_\pm$ to obtain
\be\label{asymp 2}
\Sigma_\pm  = -\log \phi_\pm +O(\phi_\pm^2)+O \left(\frac{1}{N \phi_\pm}\right).
\ee

Now we can substitute the asymptotic expansions \eqref{asymp 1} and \eqref{asymp 2} to eq. \eqref{w preprefinal}. Since the adiabaticity breaks down already for $Q/\kF=o(1)$, it is reasonable to further expand all functions in the integrand in eq. \eqref{w preprefinal} over $Q'$ and $Q''$. This finally leads to eq. \eqref{result 1}.

\section{Orthogonality catastrophe and adiabaticity\label{appendix OC and A}}

\subsection{Orthogonality catastrophe}

From eqs. \eqref{evolution of gs} and \eqref{W_Q}
 one can can calculate the orthogonality overlap
\be\label{orthogonality overlap}
\mathcal C_Q\equiv\left|\la \Phi_0 \,|\, \Phi_Q \ra \right|^2=\exp\left( - \sum_q |\alpha_Q^q-\alpha_0^q|^2  \right).
\ee
This is done with the help of the relation
\begin{align}
e^{\alpha \,  \hat a^\dagger - \overline{\alpha'} \, \hat a}
\,& \,e^{- \alpha' \, \hat a^\dagger + \overline{\alpha'} \, \hat a} \nonumber \\
= &
e^{(\alpha-\alpha') \, \hat a^\dagger}
\,\,e^{-(\alpha^*-\overline{\alpha'}) \, \hat a}
\,\,e^{-|\alpha-\alpha'|^2/2- i\, {\rm Im} \, \alpha \overline{\alpha'}}
\end{align}
valid for arbitrary $\alpha$ and $\alpha'$.

We calculate $\log \mathcal C_Q$ explicitly by substituting the expression \eqref{alpha} for $\alpha_Q^q$ in eq. \eqref{orthogonality overlap}, expanding the latter in $Q$ and performing the sum over $q$. This way  we obtain in the leading order the r.h.s. of eq. \eqref{result 0}.

\subsection{Relation between the adiabaticity and the orthogonality catastrophe}

We wish to establish that $\mathcal C_Q\simeq \mathcal F_Q$ in the limit of large $N$. The rigorous form  of this relation was proven in ref. \cite{lychkovskiy2017time}. In the considered case it reads
\be\label{accuracy}
|\mathcal C_Q- \mathcal F_Q|
\leq
\frac1{F_N}
 \int\limits_0^{Q}
dQ'
\sqrt{
\la\Psi_0|\H_{Q'}^2|\Psi_0\ra- \la\Psi_0|\H_{Q'}|\Psi_0\ra^2
}
\ee
The integral in this equation is cutoff-dependent but does not diverge with $N$. Therefore the r.h.s of the inequality can be made vanishingly  small, i.e. $o(1)$ in the limit $N\rightarrow \infty$, whenever  $Q=o(F_N)$. This way we reproduce eq. \eqref{result 0} and the first line of eq. \eqref{result 1}  without solving the  Schr\"odinger equation, with the use of a shortcut introduced in ref. \cite{lychkovskiy2017time}. The price is a superficially reduced range of validity of the results,  $Q=o(F_N)$ instead of  $Q=O(F_N)$ obtained by solving the dynamical problem.


\section{Proof that $F_N=O(N)$ is sufficient for adiabaticity\label{appendix sufficient}}
For the sake of such a proof we take the integral in eq. \eqref{beta exact} by parts. Importantly, this produces the factor $F_N/q$ which eventually does the job. This way for $q>0$ one estimates
\be\label{estimate}
|\beta_Q^q-\alpha_Q^q|^2\leq \frac{F_N^2}{2 \pi q^3 L}\, \left(A_Q^\pm\right)^2,
\ee
where
\be\label{estimate 2}
\frac{A_Q^\pm}{\kF\vs} \equiv
\left|
\frac{\partial_Q \delta^\pm_Q}{\vs-v_Q}
\right|
+
\left|
\frac{\left. \partial_Q \delta^\pm_Q\,\right|_{Q=0}}{\vs-v_0}
\right|
+
\int_0^Q
\left|\partial_{Q'}
\left(
\frac{\delta_{Q'}^\pm}{\vs-v_Q}
\right)
\right|
\,d Q'
\ee
does not depend on $q$ and is finite in the thermodynamic limit. Here the sign in $A_Q^\pm$ and $\delta^\pm_Q$ is the same as the sign of $q$.
The sum over positive $q$ in eq. \eqref{w prefinal} is bounded from above according to
\be
\sum_{q>0}  |\beta_Q^q-\alpha_Q^q|^2\leq \zeta(3)\, \left(\frac{A_Q^+\,F_N\,N}{4\pi\,\kF^2\,\vs}\right)^2,
\ee
and the sum over negative $q$ is constrained analogously. Here $\zeta(3)$ is the Riemann zeta function. This proves that whenever
\be
F_N\leq  \frac{\epsilon}{N} \frac{4\pi\, \kF^2\,\vs}{\sqrt{\zeta(3)\left( \left(A_Q^+\right)^2+ \left(A_Q^-\right)^2\right)}}
\ee
the adiabatic fidelity is bounded from below,
\be
\mathcal F_Q \geq e^{-\epsilon^2}.
\ee

\section{Integrable model: consistency check\label{appendix BA}}

Here we consider a microscopic  integrable impurity-fluid model. The model consists of $N$ fermions and a single impurity particle with a mass $m$ equal to the mass of a fermion.  Fermions do not interact with each other but couple to the impurity via the repulsive contact potential. The Hamiltonian reads
\begin{equation}\label{H McGuire}
\H_Q=\frac{(- i \, \partial_X+Q)^2}{2m}-\sum_{j=1}^{N}
\frac{1}{2m}\partial_{x_j}^2+g \sum_{j=1}^{N}  \delta(x_j-X),
\end{equation}
where $Q=F t$ is the impulse of the force,
 $X$ and $x_j$  are the coordinates of the impurity and the $j$'th fermion
respectively  and $g>0$ is the impurity-fluid coupling.

For a fixed $Q$ the model (\ref{H McGuire}) is integrable as shown by McGuire \cite{mcguire1965interacting}. In fact, this model is one of the simplest models solvable via the Bethe ansatz: Its eigenfunction can be expressed through $(N+1)\times(N+1)$ Slatter-like determinants \cite{recher2012hardcore,mathy2012quantum}. For this reason it has been possible to obtain a wealth of analytical results and to gain a number of deep insights into the physics of the model \cite{recher2012hardcore,mathy2012quantum,castella1993exact,Burovski2013,gamayun2015impurity,gamayun2016time,gamayun2018impact}. Although this model is a special case of the Yang-Gaudin model \cite{yang1967some,gaudin1967systeme}, it might deserve a separate name -- McGuire model -- due to its conceptual importance.

The integrability of the model enabled us to apply the technique of ref. \cite{lychkovskiy2017time} and relate the adiabaticity breakdown to the orthogonality catastrophe in ref. \cite{lychkovskiy2018quantum}. Here we focus on the relation between the microscopic model \eqref{H McGuire} to the effective model \eqref{H(Q)}. Our aim is to underpin the effective model \eqref{H(Q)} and better understand its validity range. In what follows the notations and conventions follow refs. \cite{gamayun2016time,gamayun2018impact}.

First, we would like to relate the effective scattering phases in eq. \eqref{H(Q)} to the microscopic scattering phases of the Bethe ansatz \cite{gamayun2018impact},
\be
\delta_{{\rm BA}\,Q}^\pm=\frac\pi2-\arctan\left(\frac{\Lambda}{\kF}\mp\frac{2\vs}{g}\right),
\ee
where $\vs$ is the Fermi velocity in the present context and the parameter $\Lambda$ can be found from the equation \cite{gamayun2018impact}
\begin{align}\label{phase}
\frac{Q}{k_F}  =
\frac{g}{2\pi\vs}
&
\bigg[
\left(\Lambda+\frac{2\vs}{g}\right)\arctan\left(\Lambda+\frac{2\vs}{g}\right)
\nonumber\\
&-\left(\Lambda-\frac{2\vs}{g}\right)\arctan\left(\Lambda-\frac{2\vs}{g}\right)
\nonumber\\
&+\frac{1}{2}\ln \frac{1+({2\vs}/{g}-\Lambda)^2}{1+({2\vs}/{g}+\Lambda)^2}\bigg].
\end{align}

To do this we consider the overlap
\be
\widetilde {\cal C}_Q \equiv \left| \langle \widetilde \Phi_Q |\Phi_Q \rangle \right|^2
\ee
between the ground state $\Phi_Q$ and the {\it noninteracting} ground state $\widetilde \Phi_Q$ of the impurity-fluid system. $\widetilde {\cal C}_Q$ can be calculated both in the microscopic model \eqref{H McGuire} \cite{castella1993exact} and the effective model \eqref{H(Q)}, with the leading order results
\be
\log \widetilde {\cal C}_Q = - \left(
\left(\frac{\delta_{{\rm BA}\,Q}^+}{\pi} \right)^2+\left(\frac{ \delta_{{\rm BA}\,Q}^-  }{\pi} \right)^2
 \right) \log N
\ee
and
\be
\log \widetilde {\cal C}_Q = - \left(
\left(\frac{\delta_Q^+}{2\pi} \right)^2+\left(\frac{ \delta_Q^-  }{2\pi} \right)^2
 \right) \log N,
\ee
respectfully. These two equations are compatible when
\be\label{identification}
\delta_Q^\pm=2 \delta_{{\rm BA}\,Q}^\pm.
\ee

Now we turn to the orthogonality overlap of interest, ${\cal C}_Q$. We have calculated it within the McGuire model in a similar manner as in ref. \cite{castella1993exact}, with the result
\begin{align}\label{orthogonality overlap BA}
\log{\cal F}_Q= &\, -\left(
\left(\frac{ \delta_{{\rm BA}\,Q}^+ - \delta_{{\rm BA}\,0}^+}{\pi} \right)^2 +\left(\frac{ \delta_{{\rm BA}\,Q}^-  - \delta_{{\rm BA}\,0}^- }{\pi} \right)^2
 \right) \nonumber \\
&
\, \times \big(\log N+O(1)\big),
\end{align}
where $O(1)$ refers to the limit of $N\rightarrow\infty$.

After accounting for eq. \eqref{identification} and expanding in small $Q$  this result agrees with  eq. \eqref{result 0}. It should be emphasised that this agreement is not limited to small $Q$ but takes place in the whole Brillouin zone, $-\kF<Q<\kF$. This indicates that the actual range of validity of the effective model can span well beyond the conservative condition~\eqref{validity range}. This agreement also suggests that the boson operators $a_q,a_q^\dagger$ in fact do not depend on $Q$ ({\it cf.} footnote \ref{footnote}), at least in the impurity-fluid system described by the microscopic Hamiltonian \eqref{H McGuire}.





\bibliography{C:/D/Work/QM/Bibs/1D,C:/D/Work/QM/Bibs/LZ_and_adiabaticity}

\end{document}